\documentclass[3p,preprint,12pt]{elsarticle}
\makeatletter\if@twocolumn\PassOptionsToPackage{switch}{lineno}\else\fi\makeatother

\usepackage{tabulary,xcolor}
\usepackage{braket}
\usepackage{amsfonts,amsmath,amssymb}
\usepackage{circuitikz}
\usetikzlibrary{angles,quotes}
\usepackage{pdflscape}
\usepackage[T1]{fontenc}
\usepackage{qcircuit}
\usepackage[doublespacing]{setspace}
\usepackage{tikz-qtree}
\makeatletter
\let\save@ps@pprintTitle\ps@pprintTitle
\def\ps@pprintTitle{\save@ps@pprintTitle\gdef\@oddfoot{\footnotesize\itshape \null\hfill}}
\def\hlinewd#1{%
  \noalign{\ifnum0=`}\fi\hrule \@height #1%
  \futurelet\reserved@a\@xhline}

\AtBeginDocument{\ifNAT@numbers \biboptions{sort&compress}\fi}
\makeatother

\usepackage{mathtools}
\usepackage{graphicx}
\usepackage{epstopdf}
\usepackage{pdflscape}
\usepackage{caption}
\usepackage{ifluatex}
\ifluatex
\usepackage{fontspec}
\defaultfontfeatures{Ligatures=TeX}
\usepackage[]{unicode-math}
\unimathsetup{math-style=TeX}
\else 
\usepackage[utf8]{inputenc}
\fi 
\ifluatex\else\usepackage{stmaryrd}\fi

\usepackage{url,multirow,morefloats,floatflt,cancel,tfrupee}
\makeatletter

\AtBeginDocument{\@ifpackageloaded{textcomp}{}{\usepackage{textcomp}}}
\makeatother
\usepackage{colortbl}
\usepackage{xcolor}
\usepackage{pifont}
\usepackage[nointegrals]{wasysym}
\makeatletter

\def\mcWidth#1{\csname TY@F#1\endcsname+\tabcolsep}

\def\cAlignHack{\rightskip\@flushglue\leftskip\@flushglue\parindent\z@\parfillskip\z@skip}
\def\rAlignHack{\rightskip\z@skip\leftskip\@flushglue \parindent\z@\parfillskip\z@skip}

\@ifundefined{etal}{}{}

\usepackage{ifxetex}
\ifxetex\else\if@twocolumn\@ifpackageloaded{stfloats}{}{\usepackage{dblfloatfix}}\fi\fi

\AtBeginDocument{
\expandafter\ifx\csname eqalign\endcsname\relax
\def\eqalign#1{\null\vcenter{\def\\{\cr}\openup\jot\m@th
  \ialign{\strut$\displaystyle{##}$\hfil&$\displaystyle{{}##}$\hfil
      \crcr#1\crcr}}\,}
\fi
}

\AtBeginDocument{%
  \@ifpackageloaded{endfloat}%
   {\renewcommand\efloat@iwrite[1]{\immediate\expandafter\protected@write\csname efloat@post#1\endcsname{}}}{\newif\ifefloat@tables}%
}%

\def\BreakURLText#1{\@tfor\brk@tempa:=#1\do{\brk@tempa\hskip0pt}}
\let\lt=<
\let\gt=>
\def\processVert{\ifmmode|\else\textbar\fi}

\@ifundefined{subparagraph}{
\def\subparagraph{\@startsection{paragraph}{5}{2\parindent}{0ex plus 0.1ex minus 0.1ex}%
{0ex}{\normalfont\small\itshape}}%
}{}

\newcommand\role[1]{\unskip}
\newcommand\aucollab[1]{\unskip}
  
\@ifundefined{tsGraphicsScaleX}{\gdef\tsGraphicsScaleX{1}}{}
\@ifundefined{tsGraphicsScaleY}{\gdef\tsGraphicsScaleY{.9}}{}
\def\checkGraphicsWidth{\ifdim\Gin@nat@width>\linewidth
	\tsGraphicsScaleX\linewidth\else\Gin@nat@width\fi}

\def\checkGraphicsHeight{\ifdim\Gin@nat@height>.9\textheight
	\tsGraphicsScaleY\textheight\else\Gin@nat@height\fi}

\def\fixFloatSize#1{}
\let\ts@includegraphics\includegraphics

\def\inlinegraphic[#1]#2{{\edef\@tempa{#1}\edef\baseline@shift{\ifx\@tempa\@empty0\else#1\fi}\edef\tempZ{\the\numexpr(\numexpr(\baseline@shift*\f@size/100))}\protect\raisebox{\tempZ pt}{\ts@includegraphics{#2}}}}

\AtBeginDocument{\def\includegraphics{\@ifnextchar[{\ts@includegraphics}{\ts@includegraphics[width=\checkGraphicsWidth,height=\checkGraphicsHeight,keepaspectratio]}}}

\DeclareMathAlphabet{\mathpzc}{OT1}{pzc}{m}{it}

\def\URL#1#2{\@ifundefined{href}{#2}{\href{#1}{#2}}}

\def\UrlOrds{\do\*\do\-\do\~\do\'\do\"\do\-}%
\g@addto@macro{\UrlBreaks}{\UrlOrds}

\edef\fntEncoding{\f@encoding}

\makeatother

\newif\ifmultipleabstract\multipleabstractfalse%
\begin{document}

\pagestyle{myheadings}

\begin{frontmatter}

\title{A joint optimization approach of parameterized quantum circuits with a tensor network}

\author[First]{Clara Ferreira Cores}

\author[First]{Kaur Kristjuhan}

\author[First]{Mark Nicholas Jones}
\ead{research@mqs.dk}


\address[First]{Molecular Quantum Solutions ApS, Maskinvej 5, 2860 Søborg, Denmark}
    
\begin{abstract}
Despite the advantage quantum computers are expected to deliver when performing simulations compared to their classical counterparts, the current noisy intermediate-scale quantum (NISQ) devices remain limited in their capabilities. The training of parameterized quantum circuits (PQCs) remains a significant practical challenge, exacerbated by the requirement of shallow circuit depth necessary for their hardware implementation.
Hybrid methods employing classical computers alongside quantum devices, such as the Variational Quantum Eigensolver (VQE), have proven useful for analyzing the capabilities of NISQ devices to solve relevant optimization problems. Still, in the simulation of complex structures involving the many-body problem in quantum mechanics, major issues remain about the representation of the system and obtaining results which clearly outperform classical computational devices.
In this research contribution we propose the use of parameterized Tensor Networks (TNs) to attempt an improved performance of the VQE algorithm. A joint approach is presented where the Hamiltonian of a system is encapsulated into a Matrix Product Operator (MPO) within a parameterized unitary TN hereby splitting up the optimization task between the TN and the VQE. We show that the hybrid TN-VQE implementation improves the convergence of the algorithm in comparison to optimizing randomly-initialized quantum circuits via VQE.
\end{abstract}


\end{frontmatter}

\section{Introduction}
Quantum computing has gained attention for its unique potential for handling complex problems that classical computers struggle to solve. Specific problems within physics and chemistry have a high computational cost which renders classical methods intractable for specific cases \citep{RevModPhys.92.015003}. Quantum algorithms offer a possibility to obtain accurate solutions with a computational speed-up. This is of great interest for the quantum chemistry field, where accurate models (e.g. CCSD(T)) require high computational efforts to reach correct solutions and are not suitable for high-throughput screening of a large molecular data set \citep{doi:10.1021/acs.jctc.3c00419}.
\\
Hybrid algorithms, which combine classical processing units (e.g. CPUs, GPUs) with quantum processing units (QPUs), have shown to be suitable at solving problems on existing quantum hardware \citep{chan2023hybrid}.
\\
One well-known algorithm is the Variational Quantum Eigensolver (VQE), which determines the ground state of a given Hamiltonian \citep{2022PhR...986....1T}. The VQE algorithm does this by searching for a quantum state that minimizes the expectation value of the Hamiltonian. States are prepared and evaluated on a quantum computer with parameterized quantum circuits (PQCs). The values of the parameters are optimized with classical optimization routines such as gradient descent. This approach has the advantage of keeping the quantum circuit depth shallow and hence mitigating noise in NISQ devices. Despite their apparent suitability for NISQ devices, they still face important challenges involving their trainability, accuracy, and efficiency \citep{2020arXiv201209265C}.
\\
The performance of a VQE is heavily influenced by the particular choice of PQC and its initial parameter values in relation to the problem that the VQE is attempting to solve. The structure of the PQC determines the optimization landscape, which may contain undesirable features, such as barren plateaus \citep{2018arXiv181201041Z}.
\\
A barren plateau is an area of the optimization landscape, where the gradients of all parameters are close to zero, for an extended region of the parameter space. Such an area is not an optimization minimum, but it causes optimization algorithms to stall or terminate because true minima also have vanishing gradients. The vanishing of gradients has been observed both for shallow \citep{2021NatCo..12.1791C} and deep PQCs \citep{2018NatCo...9.4812M} that are randomly initiliazed.
\\
Even if the optimization landscape contains barren plateaus, it is possible to avoid them by choosing a good set of initial parameters. Recent work reported in \cite{rudolph2023synergistic} has demonstrated the possibility of using tensor network (TN) based classical techniques to improve the performance of PQCs by finding suitable initial parameters.
In fact, the importance of refining initial states through classical methods has been further highlighted in \cite{fomichev2024initial}. Their use of advanced techniques for state preparation, including the representation of the ground-state wavefunction using TN methods, led to higher probabilities of low-energy estimates, resulting in fewer repetitions. A similar improved outcome can be observed in the average performance of our TN-VQE algorithm compared to that of the standard VQE.
In this research contribution, we will explore a complementary approach which uses a unitary tensor network (uTN) to aid a PQC during the optimization procedure of VQE. TN methods have become widely popular in numerical simulations algorithms, expanding the horizons of classical simulability \citep{rudolph2023synergistic}. Their way of efficiently representing the structure of entanglement in quantum many-body systems has placed them at the forefront of simulations within condensed-matter physics \citep{Haghshenas_2021}.
\\
The efficiency of TN methods to assist a VQE algorithm is put to the test, comparing the performance of the TN-VQE with a purely VQE algorithm tackling the same Hamiltonian, a 1D transverse field Ising model.
\\
The structure of this chapter is as follows: the next section will cover the theory of the mathematical concepts applied for the simulation of the Hamiltonian. In Section \ref{sec:tn-vqe} the building blocks of the TN-VQE algorithm are explained. The methodology steps are presented in Section \ref{sec:methodology} and some notes on implementation, consistency checks and validation are provided in Section \ref{sec:implementation}.
The results obtained, along with a comparison to randomly initialized quantum circuits, are exhibited in Section \ref{sec:results}, and concluding remarks are given in Section \ref{sec:conclusion}.
\\
\section{Theoretical Background}
\subsection{Tensor Networks (TN)}
\begin{figure}[htbp!]
    \centering
    \includegraphics[scale=0.4]{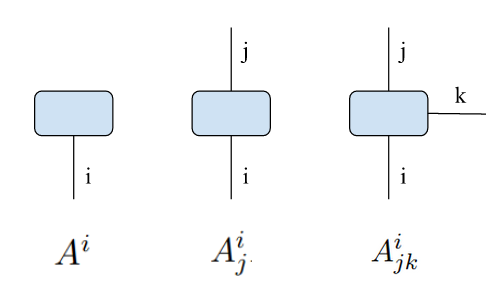}
    \caption{Diagram notation for tensors. From left to right: Rank 1 tensor, a vector; Rank 2 tensor, a matrix; Rank 3 tensor.}
    \label{fig:tensors}
\end{figure}

\begin{figure}[htbp!]
    \centering
    \includegraphics[scale=0.4]{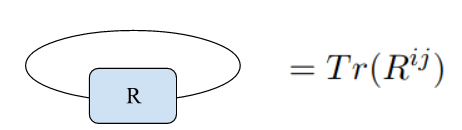}
    \caption{Diagram representation of the trace operation on a tensor.}
    \label{fig:trace}
\end{figure}

\begin{figure}[htbp!]
    \centering
    \includegraphics[scale=0.45]{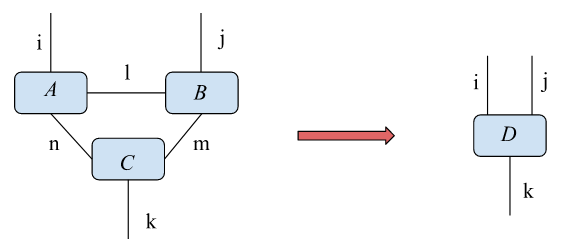}
    \caption{Diagram representation of the contraction of a tensor network, resulting in tensor $D^{ij}_k$.}
    \label{fig:network_contraction}
\end{figure}

A tensor is a multidimensional array of complex numbers. The rank of a tensor refers to its number of dimension, each represented with an index. Tensors can be represented by diagrams, in which the number of lines corresponds to their rank as seen in Figure \ref{fig:tensors}.
\\
Possible operations between tensors include tensor products, tensor contraction, and the decomposition of a tensor into lower rank structures (e.g. SVD) \citep{TNnet}. For instance, a matrix multiplication (tensors of rank 2) can be thought as a tensor product followed by a contraction over the shared index:
\begin{equation}
    A^i_{j} = \sum_k B^i_k C^k_j
\end{equation}
\\
An index contraction is the sum over all the possible values of the repeated indices of a set of tensors. Tensor contraction removes a line in a tensor diagram and can only be performed if both ends of the line terminate at a tensor. A contraction in which both ends of a line connect to the same tensor is called a trace (Figure \ref{fig:trace}).
\\
A group of tensors, which may be connected by various lines, where some, or all, of its indices are contracted according to some pattern is called a Tensor network (TN) \citep{Or_s_2014}. The lines connecting one tensor to another in a network are called virtual bonds. Performing possible tensor contractions in a tensor network is called contracting the network, as in the example shown in Figure \ref{fig:network_contraction} \citep{video_TN}.
\\
Quantum states can be represented in terms of networks of interconnected tensors, which in turn capture the relevant entanglement properties of a system. The use of TN patterns can be accelerated with classical computational resources such as graphical and tensor processing units (GPUs and TPUs), raising the performance of the state-of-the-art classical algorithms \citep{rudolph2023synergistic, Bridgeman_2017}.
\\
The simple way of representing complex equations with tensor networks has been applied to many use-cases and areas of research: e.g. classification of novel phases of matter \citep{PhysRevB.81.064439, PhysRevB.84.165139}, machine learning \citep{2017arXiv170809165C, 2022arXiv220702851S}, and quantum computation \citep{NIPS2016_5314b967, Pan_2020}.
TNs are also of special interest for performing calculations on noisy intermediate-scale quantum (NISQ) devices given that TNs can decompose high-rank linear transformations (expressed as tensors with many indices) into a product of multiple lower rank tensors, each of which contains only a relatively small number of parameters \citep{TNnet}.
By introducing a parameterized TN, one can perform unitary transformations on the Hamiltonian with a classical computer, which would otherwise be applied to the quantum state on a quantum computer.

\subsection{The Variational Quantum Eigensolver (VQE)}
The VQE method aims to find the ground state energy of an Hamiltonian. To achieve this, the algorithm optimizes a parameterized quantum circuit (PQC) with respect to a cost function where the circuit consists of unitary quantum gates \citep{chan2023hybrid}.
\\
The optimization landscape is defined by a cost (loss) function:
\begin{equation}
E(\phi) = \braket{\psi_0 | \hat{U}^{\dagger}(\phi)\hat{H}\hat{U}(\phi) | \psi_0}
\end{equation}
where $\hat{H}$ is the problem Hamiltonian, $\ket{\psi_0}$ is the initial state and $\hat{U}(\phi)$ is an unitary operator, depending on a set of angle parameters $\phi$.
\\
The PQC prepares the state $\hat{U}(\phi)\ket{\psi_0}$ and applies the transformation $\hat{U}(\phi)$. This circuit is called the Ansatz circuit. The choice of the initial state, as well as the functional form of $\hat{U}(\phi)$ influences whether a barren plateau is encountered during optimization.
\\
When choosing an Ansatz, a balance is needed between the accurate representation of the Hamiltonian and the difficulty of its implementation in the current NISQ devices. There are for instance hardware efficient Ansätze, as the one used in superconducting arrays, where layers of single qubit gates are coupled with native entangling unitaries \citep{Kandala_2017}. Whereas other approaches focus on representing more accurately the Hamiltonian in question, e.g. coupled cluster methods \citep{10.1063/1.5141835, 2021arXiv210802792S}.
\\
The attempt in this research study is to test if the TN-VQE approach can be a solution to this dilemma, by better encapsulating the complexity of quantum many-body systems within shallow circuit depth.

\subsection{The Quantum Many-body Problem}
\begin{figure}[htbp!]
    \centering
    \includegraphics[scale=0.3]{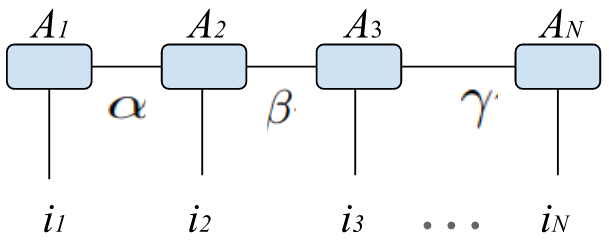}
    \caption{Graphical representation of the MPS ansatz for the many-body wavefunction. The $i_n$ stands for each particle's local Hilbert space of dimensionality $d$. Meanwhile the greek letters $\alpha, \beta, ...$ are used to name the bond indices. Thus, $A_i$ stands for tensors of dimension $m \times m \times d$, mapping between the bonds of dimension $m$ and the $d$ dimensional local Hilbert space.}
    \label{fig:MPS}
\end{figure}

Representing quantum many-body systems is a challenging task. The more particles a system contains, the more complex the interactions become. The many-body Hilbert space is constructed as the tensor product of the single particle Hilbert spaces:
\begin{equation}
    \mathcal{H}=\bigotimes_{i=1}^N\mathcal{H}^i
\end{equation}
where $\mathcal{H}^i$ is the individual Hilbert space for each particle $i$. If the dimension of each individual Hilbert space is $\dim(\mathcal{H}^i)= d$, then the dimension of the many-body Hilbert space is $\dim(\mathcal{H})=d^N$. Hence, the addition of particles to the system results on the exponential growth of the Hilbert space we ought to simulate.
\\
One way around this issue is to work with a compressed representation of the system, which is exactly what TNs provide. At low-dimensions, TNs have been successful in encoding the ground state of many-body quantum systems \citep{Nagy_2011}. The wavefunction of a many-body quantum system can be expanded in a basis as:
\begin{equation}\label{eq:wavefunction}
    \ket{\psi} = \sum_{i_1 , i_2...i_N} c^{i_1,i_2...i_N} \ket{i_1,i_2...i_N}
\end{equation}
where $i_n$ denotes the quantum numbers associated to the basis state $\ket{i_n}$. If we represent the $c^{i_1,i_2...i_N}$ coefficients as a contraction of tensors, then the exponential dependence on the number of required parameters of the system size can be reduced to a polynomial dependence \citep{course_Cologne}.
\\
For systems with one spatial dimension, the simplest tensor network structure takes form of a matrix product state (MPS) \citep{Dalzell_2019}. In this form Hamiltonians describe only local interactions on a spin chain and can be generalized to higher dimensions. One can perform a singular value decomposition (SVD), which leads to less parameters to represent the quantum state \citep{course_Cologne}. The graphical representation of the MPS Ansatz for the many-body wavefunction can be seen in Figure \ref{fig:MPS}.
\\
Every one of the tensors $A_i$ in Figure \ref{fig:MPS} can be seen as a mapping between the $m \times m$ Hilbert space from their virtual bonds to the $d$ dimensional local Hilbert space of the outward leg, the physical leg to take measurements from. For example, the mapping provided by tensor $A_2$ can be expressed as:
\begin{equation}\label{eq:Mapping_A}
    \mathcal{M}_{(A_2)} = \sum^d _{i=2}\sum^m _{\alpha, \beta =1} A_2^{\alpha,\beta} \ket{i_2} \bra{\alpha,\beta}
\end{equation}
where the local Hilbert space $\ket{i_2}$ has the dimension $d$ and $\bra{\alpha,\beta}$ is of dimension $m \times m$.
\\
From the diagram in Figure \ref{fig:MPS}, the MPS Ansatz for the many-body wavefunction can be written as:
\begin{equation}\label{eq:MPS_ansatz}
    \ket{\psi} =\sum _{i_1, i_2...i_N}\sum_{\alpha, \beta, ...} A_1 [i_1]_{\alpha} A_2 [i_2]_{\alpha,\beta}...A_N [i_N]_{\gamma} \ket{i_1 i_2 ... i_N}
\end{equation}
where $[i_n]$ are the basis states with their indices referring to the bond variables. The sum over the bond variables $\alpha, \beta, ...$, for each $A_n [i_n]_{\alpha,\beta}$ can be seen as a matrix when being contracted. Therefore, the resulting expression for the wavefunction is of the form:

\begin{equation}\label{eq:MPS}
    \ket{\psi} = \sum_{i_1 , i_2...i_N} A^{i_1}A^{i_2}...A^{i_N} \ket{i_1 i_2...i_N}
\end{equation}
where $A^{i_j}$ are the local matrices describing each particle. Note that $A^{i_1}$ and $A^{i_N}$ are row and column vectors. The bond indices corresponding to the inner connections between the matrices have been omitted in the equation above. This is the reason why there is a single index per tensor in the network.
Compared with the previous wavefunction from equation \ref{eq:wavefunction}, the parameters describing the many-body system have been reduced from $d^N$ to $N\cdot d \cdot m^2$.

\begin{figure}[htbp!]
    \centering
    \includegraphics[scale=0.5]{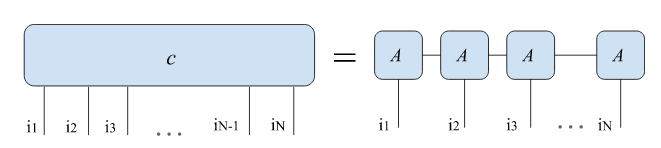}
    \caption{Decomposition of the many-body wavefunction into a MPS tensor network.}
    \label{fig:MPS_decomposition}
\end{figure}

\begin{figure}[htbp!]
    \centering
    \includegraphics[scale=0.52]{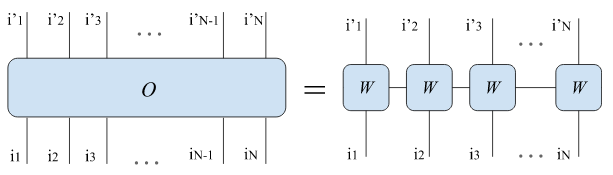}
    \caption{Decomposition of an operator acting on a many-body system into a MPO structure composed by $W^{i_n, i'^n}$ tensors.}
    \label{fig:MPO_decomposition}
\end{figure}

An operator acting on the many-body system can be written as:
\begin{equation}
    \bra{i'_1 i'_2...i'_N} \hat{O} \ket{i_1 i_2...i_N} = O^{i'_1 i'_2...i'_N}_{i_1 i_2...i_N}
\end{equation}
Once again the indices of the tensor $O$, which can be seen graphically in Figure \ref{fig:MPO_decomposition}, represent the open legs.
This higher rank tensor can be decomposed in what is called a Matrix Product Operator (MPO), similarly to the approach used to decompose the wavefunction into an MPS, represented in Figure \ref{fig:MPS_decomposition}. The resulting expression is:
\begin{equation}\label{eq:MPO_OBC}
    O = \sum_{i_1 , i_2...i_N}\sum_{i'^1 , i'^2...i'^N} W^{i_1, i'1} W^{i_2,i'2}...W^{i_N,i'N} \ket{i_1 i_2...i_N} \bra{i'_1 i'_2...i'_N}
\end{equation}

The tensors at the edges, $W^{i_1, i'1}$ and $W^{i_N,i'N}$ represent the row and column vectors, but with an additional dimension than in the MPS case, as visible in Figure \ref{fig:MPO_decomposition}.
MPS and MPOs are easily implemented within the density matrix renormalization group (DMRG) computational framework, which has become a popular toolkit within the quantum chemistry field \citep{chan2016matrix}.
\\
In this research contribution, the system studied is a finite 1D spin-chain with only nearest-neighbour interactions.
Despite the apparent simplicity of the system, one-dimensional spin chains have provided great insights on complex many-body problems \citep{LCKwek_2009}.

\section{TN-VQE}\label{sec:tn-vqe}
The TN-VQE algorithm studied in this work is inspired by the research conducted in \cite{2022arXiv221210421H} where a minimization problem to reach the ground state energy $E_g$ of the Hamiltonian $\hat{H}$ was solved by applying a hybrid approach of tensor networks and the VQE. The minimization problem can be expressed as:
\begin{equation}\label{eq:min_theta_phi}
    E_g=\min_{\theta, \phi} \bra{0} U^{\dagger}(\phi)U^{\dagger}(\theta) HU(\theta)U(\phi) \ket{0}
\end{equation}
where the TN ansatz is parameterized by the vector $\theta$, be that a single scalar or an array of values. Meanwhile the quantum circuit is parameterized by $\phi$. Thus, if $\theta$ is frozen, the minimization is performed as a traditional VQE, while if $\phi$ is frozen, it turns into a classical parameter optimization method. Attaching unitary tensor networks ($U(\theta)$) around a MPO Hamiltonian can significantly enhance its representation and improve the accuracy of energy value estimations \citep{Haghshenas_2021}.
\\
The parameterization of the TN is adapted to the 1D transverse field Ising model. Two different parameterizations are considered to test how the number of parameters affects the success of the TN ansatz.
In our research contribution, the method simultaneously updates both the $\theta$ and $\phi$ parameters within the same cost function.
In the following subsections more in-depth details about some individual steps of the algorithm are discussed before presenting the methodology applied in our research contribution.
\\
\subsection{Pauli Strings}\label{sec:pauli_strings}
To run VQE on a gate-based quantum device, the problem Hamiltonian is expressed in terms of Pauli operators. Given a general Hamiltonian $H$ written as a $2^n \times 2^n$ matrix, the Pauli decomposition will lead to:
\begin{equation}\label{eq:Pauli_decomposition}
  \begin{array}{l}
    H = \sum_x c_{P_x} P(x) \\
    {}\\

    P(x) = (\sigma_{x_{n-1}} \otimes ... \otimes \sigma_{x_0} )
  \end{array}
\end{equation}
where $\sigma_i$ are the Pauli matrices, $\sigma_{\{0, 1, 2, 3\}} = \{I, X, Y, Z\}$, given an input string $x = x_{n-1} ... x_0 \in \{0, 1, 2, 3\}^n$. $c_{P_x}$ are the Pauli coefficients obtained from the orthogonal projection of the system:
\begin{equation}
    c_{P_x} = \frac{1}{2^n}Tr[P(x)H]
\end{equation}
which can be computed with the respective trace over the TN structure \citep{romero2023paulicomposer}.
This allows for the measurements to be performed in a two-level basis, as required by the quantum devices. Each Pauli matrix is operating on a single-qubit, and each of the possible tensor product combinations for the selected number of qubits is called a Pauli string. To obtain the energy values, one needs to perform the weighted sum over the expectation values of all Pauli strings $\braket{P(x)}$ occurring in the Hamiltonian:
\begin{equation}
\braket{H} = \sum_{x} c_{P_{x}} \braket{P(x)}
\end{equation}
where $\braket{H}$ is the expectation value of the Hamiltonian.

The expectation values for each of the Pauli strings are determined independently, repeating the measurement of the TN structure for each of the possible combinations.

\subsection{Hamiltonians}\label{sec:Hamiltonians}
The 1D Transverse-Field Ising Model is a quantum version of the classical Ising Model, which features a lattice with nearest neighbour interaction and a perpendicular external magnetic field. The model can be described by the Hamiltonian:
\begin{equation}\label{eq:H_1D}
H = - J \left( \sum_{i,j} \sigma^z_i \sigma^z_j - g \sigma^x_i \right)
\end{equation}
where the lattice is along the $z$-axis, while the orthogonal magnetic field is on the $x$-axis. Hence, the interactions between neighbours are represented by the spin projections along the $z$-axis, meaning the corresponding Pauli matrices considering the decomposition of the Hamiltonian based on Equation (\ref{eq:Pauli_decomposition}). $J$ is a pre-factor with units of energy, and $g$ is a coupling coefficient representing the relative strength of the external magnetic field compared to the nearest neighbour interaction.
\\
Following the procedure described to decompose quantum operators into MPOs structures (Equation (\ref{eq:MPO_OBC})) the Hamiltonian above can be turned into a MPO formed by the product of tensors $W^{[i]}$.
\\
The tensors corresponding to the decomposition of the Hamiltonian in Equation (\ref{eq:H_1D}) are $3\times 3$ matrices with $2\times 2$ matrices as elements. The first and last tensors of the lattice lack one dimension compared to the rest, being row and column vectors respectively. Thus, the resulting matrices are:

\begin{equation}
W^{[1]} = \begin{pmatrix}I_1 & -\sigma_z^1 & g \sigma^x_1\end{pmatrix}
\end{equation}
the initial tensor corresponding to a row vector, with dimension $3\times 2\times 2$,

\begin{equation}
W^{[i]} = \begin{pmatrix}
I_i & -\sigma_z^i & g \sigma_x^i \\
0 & 0 & \sigma_z^i \\
0 & 0 & I_i \\
\end{pmatrix}
\end{equation}
the intermediate tensors with dimensions $3\times 3 \times 2 \times 2$,

\begin{equation}
W^{[N]} = \begin{pmatrix}I_N & -\sigma_z^N & g \sigma^x_N\end{pmatrix}^T
\end{equation}
and the last last tensor in the MPO chain corresponding to a column vector with dimensions $3\times2\times2$.
\\
The \(I_i\), \(\sigma_i^x\), \(\sigma_i^y\), and \(\sigma_i^z\) are the identity, Pauli-X, Pauli-Y and Pauli-Z operators acting on the i-th qubit \citep{2022arXiv221210421H}. The dimensions of the open legs (i.e. physical legs) in the MPO structure is equal to 2, fitting the two-level structure of a quantum device. Higher dimensions can be kept in the virtual bonds between the tensors.
\\
\subsection{Conditions for selecting TN U(\(\theta\))}\label{sec:uTN_1}
\begin{figure}[htbp!]
    \centering
    \includegraphics[scale=0.4]{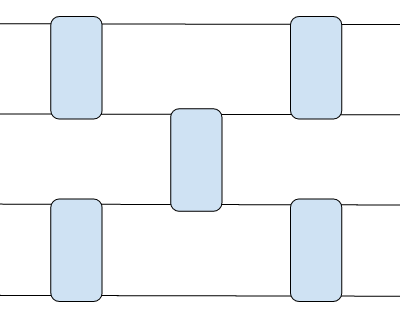}
    \caption{Structure of the unitary MPO attached at each end of the Hamiltonian ($U^{\dagger}(\theta)HU(\theta)$). Each box is a unitary MPO obtained from Equation \ref{eq:single_mpo_1}, $M^{ijkl}(\theta)$ in the case of $U(\theta)$ and $M^{\dagger jilk}(\theta)$ for the $U^{\dagger}(\theta)$ network. The dimensions of each physical input and output is of order 2.}
    \label{fig:U_theta}
\end{figure}

In order to construct unitary tensors, unitary matrices must be selected accordingly. An arbitrary rotation matrix parameterized by three angles can be written down as follows:

\begin{equation}\label{eq:arbitrary_matrix}
R(\theta_1,\theta_2,\theta_3)=
\begin{bmatrix}
    e^{\frac{i(\theta_1 + \theta_3)}{2}} \cdot \cos{\frac{\theta_2}{2}} & e^{-\frac{i(\theta_1 - \theta_3)}{2}} \cdot \sin{\frac{\theta_2}{2}} \\
    e^{-\frac{i(\theta_1 - \theta_3)}{2}} \cdot \sin{\frac{\theta_2}{2}}      & e^{\frac{i(\theta_1 + \theta_3)}{2}} \cdot \cos{\frac{\theta_2}{2}}
\end{bmatrix}
\end{equation}
where $\theta_1,\theta_2,\theta_3$ are the parameters to optimize.
\\
The use of more parameters provides greater expressivity, whereas less parameters makes the optimization procedure faster. To evaluate how efficiency and accuracy depend on the number of parameters, a unitary tensor is constructed in parallel to the previous one. The chosen matrix for this second construction is a special case of the previous matrix, Equation (\ref{eq:arbitrary_matrix}), which is known as the phase shift gate matrix and is parameterized by a single angle:
\begin{equation}\label{eq:single_parameter_matrix}
R(\theta_1)=
\begin{bmatrix}
    e^{-i\theta_1}  & 0 \\
    0   & e^{i\theta_1}
\end{bmatrix}
\end{equation}
\\
Both matrices in Equation \ref{eq:arbitrary_matrix} and Equation \ref{eq:single_parameter_matrix} are used independently to construct unitary tensors.
\begin{figure}[htbp!]
    \centering
    \includegraphics[scale=0.5]{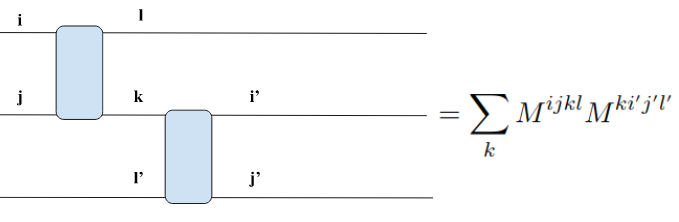}
    \caption{Einstein summation of the contraction for each connecting leg between tensors in Figure \ref{fig:U_theta}. This results in 6 contractions of this type to obtain the parameterized contracted unitary TN $U(\theta)$.}\label{fig:leg_contraction}
\end{figure}
One way to construct a unitary tensor network (uTN), is to use unitary tensors as building blocks \citep{2022arXiv221210421H, Haghshenas_2021}. The uTN is composed of local unitary tensors $u$:
\begin{equation}
    u^{\dagger}u= uu^{\dagger}= \mathcal{I}
\end{equation}
These unitary tensors are obtained from the previous matrices as follows:
\begin{equation}\label{eq:single_mpo_1}
u = M^{ijkl}(\theta) = R(\theta)^{ij} \otimes R(\theta)^{kl}
\end{equation}
where $\theta$ is a vector which depends on the matrix of choice and contains more or less parameters accordingly.
\\
$R(\theta)^{ij}$ is the parameterized matrix, and the upper indices are added to further point out the increment in dimensions (legs) which originate from the tensor product of the matrices. The change in notation of the local tensor from $u$ to $M^{ijkl}(\theta)$ has been chosen to represent more clearly the number of legs and the parameterization of the uTN structure.
\\
The tensors $M^{ijkl}(\theta)$ are then connected through a MPO network structure to build $U(\theta)$. This is achieved through tensor contraction, as illustrated in Figure \ref{fig:leg_contraction}.
The resulting network pattern for $U(\theta)$ is shown in Figure \ref{fig:U_theta}.
\begin{figure}[htbp!]
    \centering
    \includegraphics[scale=0.5]{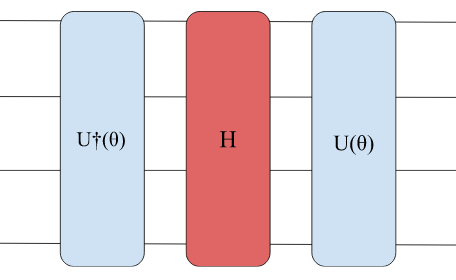}
    \caption{TN representation of the problem Hamiltonian by the use of unitary networks $U(\theta)$ and $U^{\dagger}(\theta)$, to meet the requirement of sharing the same ground state energy, as stated in Equation (\ref{eq:UHU}) where $H=H_{MPO}$.}
    \label{fig:UHU}
\end{figure}

\subsection{Unitary MPO}\label{sec:uTN_2}
The $U(\theta)$ structure is a MPO of the dimension $2 \times 2 \times 2 \times 2$. These are the desired dimensions to be assembled in a symmetric network that will have as many physical input and output legs as the Hamiltonian size requires, while keeping the bond dimension at 2.  To check that this MPO structure is indeed a unitary TN, the similarity transformation is used as a test:
\begin{equation}\label{eq:UHU}
H_{TN}= U^{\dagger}(\theta)H_{MPO}U(\theta)
\end{equation}
where  $U(\theta)$ is the network structure in Figure \ref{fig:U_theta} and $H_{MPO}$ is the Hamiltonian referred to in Section \ref{sec:Hamiltonians}. The resulting $H_{TN}$ is depicted in Figure \ref{fig:UHU}. Unitary transformations preserve the of a matrix, thus the eigenvalues of $H_{TN}$ must have the same eigenvalues of $H_{MPO}$ for the transformation being unitary ($U(\theta)$).

The procedure to construct $U^{\dagger}$ is the same as for $U(\theta)$, only that the network relies on  $(M^{jilk})^*(\theta)$ as tensors.

\subsection{Optimization strategies}\label{sec:optimization}
\begin{figure}[htbp!]
    \centering
    \includegraphics[scale=0.53]{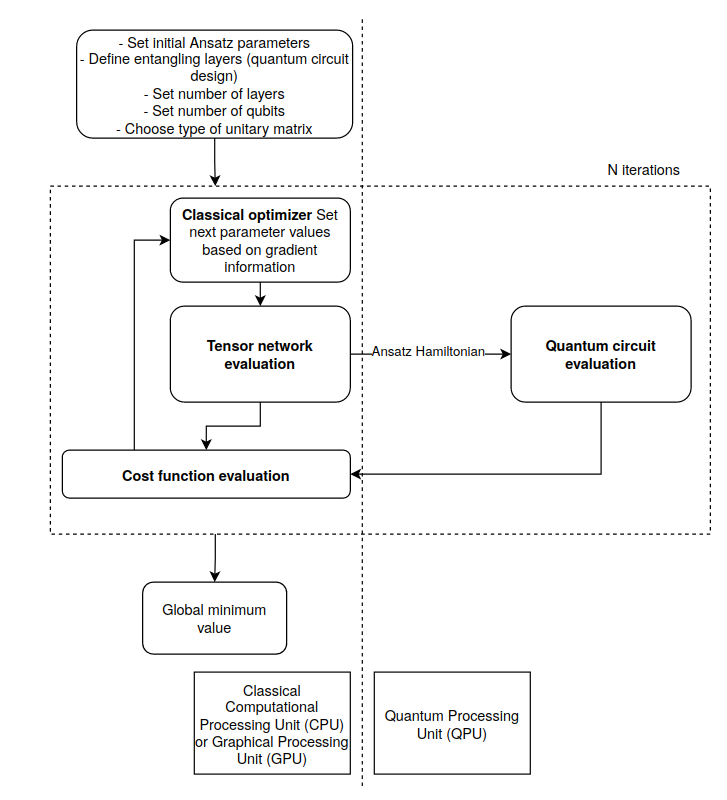}
    \caption{Diagram showcasing the optimization pipeline of the TN-VQE algorithm.}
    \label{fig:optimization_diagram}
\end{figure}

The joint approach of the TN-VQE method has two parts to optimize: the parameterized TN and the quantum circuit.
Their respective sets of parameters, $\theta$ and $\phi$, are optimized simultaneously by minimizing the cost function given in equation \ref{eq:min_theta_phi}.
\\
After each parameter update the quantum optimization is performed by performing the measurements of the quantum circuit on the quantum device simulator. The expectation values from this simulation are employed in the cost function evaluation, and repeated through enough iterations for the convergence criterion to be met (see Figure \ref{fig:optimization_diagram}).

\newpage

\section{Methodology}\label{sec:methodology}

The general methodology steps of this research work are presented in summarized form, along with references to previous sections with more specific details on the topics.

\begin{enumerate}
\item[]{\textbf{1. Formulate the problem Hamiltonian for a given system}}
\\
A one-dimensional system based on the Transverse-Field Ising Model is defined. It is composed by a finite spin chain with nearest neighbor interactions and an external perpendicular magnetic field.
\\
(see Section \ref{sec:Hamiltonians})
\\
\item[]{\bf{2. Translate Hamiltonian from Pauli matrices to MPO representation}}
\\
The quantum operator is expressed as a MPO (Equation (\ref{eq:MPO_OBC})):
\begin{equation}
H \approx H_{MPO} = J \prod_i^n W^{[i]}
\end{equation}
(see Section \ref{sec:Hamiltonians})
\\
\item[]{\textbf{3. Construct unitary TN U($\theta$)}}
\\
The final TN structure is a unitary MPO, tested through the similarity transformation of Equation (\ref{eq:UHU}). Two parallel $U(\theta)$ constructions are generated and checked for their consistency. Their corresponding unitary matrices are defined by Equation (\ref{eq:arbitrary_matrix}) and Equation (\ref{eq:single_parameter_matrix}).
\\
The construction of the local tensors encompassing the uTN structure is of the form:
\begin{equation}\label{eq:single_mpo_2}
M^{ijkl}(\theta) = R(\theta)^{ij} \otimes R(\theta)^{kl}
\end{equation}
where $R(\theta)$ is the unitary matrix of choice. Then the local tensors $M^{ijkl}(\theta)$ are connected following the pattern provided in Figure \ref{fig:U_theta}. This structure can be described mathematically as:
\begin{equation}\label{eq:add_local_tensors_1}
U(\theta)= \sum_{befijm} M^{abde}_{0}(\theta_{0}) M^{befg}_{1}(\theta_{1}) M^{efij}_{2}(\theta_{2}) M^{hilm}_{3}(\theta_{3}) M^{jklm}_{4}(\theta_{4})
\end{equation}
where the sum over the shared indices ($b,e,f,i,j,m$) is the contraction to generate the uTN structure ($U(\theta)$) with $\theta = (\theta_{0}, \theta_{1}, \theta_{2}, \theta_{3}, \theta_{4})$.
\\
(see Sections (\ref{sec:uTN_1}) and (\ref{sec:uTN_2}))
\\
\item[]{\textbf{4. Combine Hamiltonian and U($\theta$) to generate TN}}
\\
Three different tensor networks have been built up to this this point: $U(\theta)$, $H_{MPO}$ and $U^{\dagger}(\theta)$, and they have to be connected to obtain $H_{TN}$ (Equation (\ref{eq:UHU})). The diagram representation of the resultant $H_{TN}$ can be seen in Figure \ref{fig:UHU}. The contraction between the structures can be expressed as follows:
\begin{equation}\label{eq:add_local_tensors_2}
H_{TN}^{\alpha, \omega} = \sum_{\beta, \gamma} U^{\dagger}(\theta)^{\alpha, \beta} H_{MPO}^{\beta, \gamma} U(\theta)^{\gamma, \omega}
\end{equation}
here each of the parameters $\alpha, \beta, \gamma$ and $\omega$ represent a set of 4 indices ($i,j,k,l$), meaning four open legs, since the TN structures contracted above are of rank 8.
Contracting the three structures results in a TN of rank 8 (8 open legs).
\\
(see Section \ref{sec:uTN_2})
\\
\item[]{\textbf{5. Contract TN to calculate traces for Pauli coefficients}}
\\
In order to extract the expectation values from the $H_{TN}$ structure, the measurements are taken in the Pauli basis. Given the Pauli decomposition of the Hamiltonian (Equation (\ref{eq:Pauli_decomposition})), the energy values can be computed as follows:
\begin{equation} \label{eq:lin_comb_exp_val}
    E(\theta, \phi)= \sum_P c_P (\theta) \langle P \rangle_{\phi}
\end{equation}
\\
The coefficients $c_P (\theta)$ of equation (\ref{eq:lin_comb_exp_val}) are obtained by tracing the TN structure with each of the possible Pauli string combinations:
\begin{equation}
c_P (\theta)= \frac{1}{2^4}Tr[ P U^{\dagger}(\theta)H U(\theta)]
\end{equation}\label{eq:cp}
(see Section \ref{sec:pauli_strings})
\\
\item[]{\textbf{6. Define quantum circuit U($\phi$)}}
\\
The circuit chosen is a strongly entangled layer circuit, consisting of layers of single qubit rotations and entanglers; inspired by the circuit-centric classifier from \cite{Schuld_2020}. The weights for each layer are randomized, and the number of layers are adjusted according to the desired output.
\\
\item[]{\textbf{7. Optimize the parameters $\theta$ and $\phi$ with VQE}}
\\
The minimization problem to be solved is described in Equation (\ref{eq:min_theta_phi}).
Rather than optimizing each set of parameters separately, the optimization in the TN-VQE approach is done simultaneously. Thus, both $\theta$ and $\phi$ are obtained from the same cost function with a gradient descent optimizer.
\\
(see Section \ref{sec:optimization})
\end{enumerate}

\section{Implementation, Consistency Checks and Validation}\label{sec:implementation}
The whole algorithm has been implemented in the Python programming language version 3.8.
The package used for constructing and contracting the tensor network structures is TensorNetwork \citep{roberts2019tensornetwork}.
The quantum circuit and the VQE method (classical optimizer) has been implemented with the Pennylane package \citep{bergholm2022pennylane}.
The parameterized circuit of choice is the StronglyEntanglingLayers class in the Pennylane package.
\\
Consistency checks have been implemented as a collection of tests to assure the network is unitary.
The tests are:
\begin{itemize}
    \item The trace of the unitary MPO with its conjugate transpose counterpart being equal to the identity matrix:
\begin{equation}
    Tr(U^{\dagger}(\theta) U(\theta)) = \mathcal{I}
\end{equation}
    \item The eigenvalues obtained from the $H_{MPO}$ being equal with those of the $H_{TN}$.
    \\
    \item The eigenvalues obtained from the $H_{MPO}$ being equal to the expectation values obtained when diagonalizing the Hamiltonian constructed in $Pennylane$, when using the same nonzero coefficients from the Pauli strings.
\end{itemize}

The consistency checks are enforced throughout the simulations to ensure that no unexpected deviations from mathematical constraints occur from implementation errors.

\section{Results}\label{sec:results}
\begin{figure}[htbp!]
    \centering
    \includegraphics[width=\linewidth]{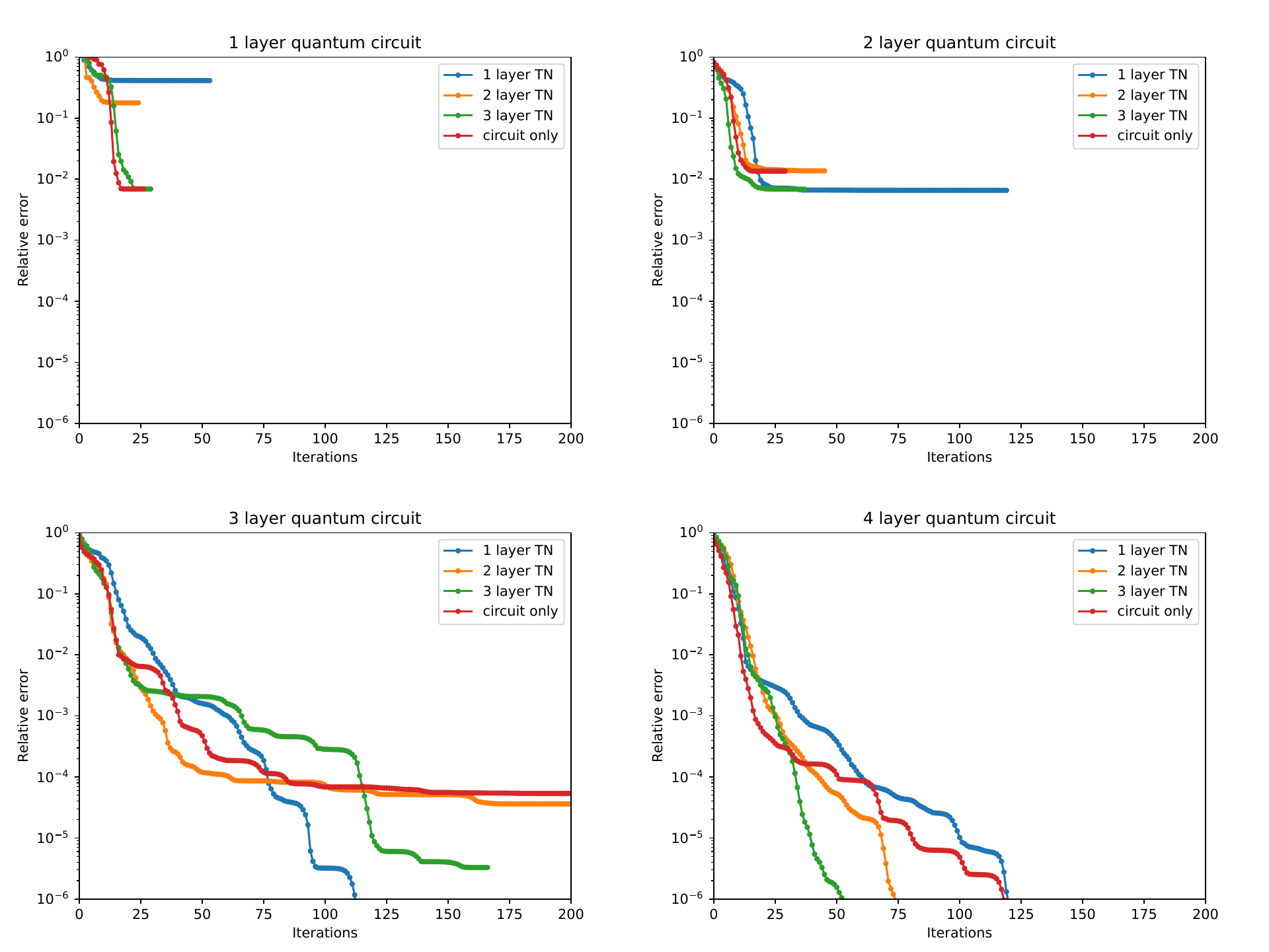}
    \caption{Relative error of the ground state energy of the transverse-field Ising model as a function of iterations in the VQE optimization procedure. The number of layers in the quantum circuits increases from 1 to 4 reading the plots from left to right. All tensor networks are constructed from tensors with one parameter. Lines which terminate before 200 iterations do so when the optimizer is unable to update the parameters due to a vanishing gradient.}
    \label{fig:single_parameter_results}
\end{figure}

\begin{figure}[htbp!]
    \centering
    \includegraphics[width=\linewidth]{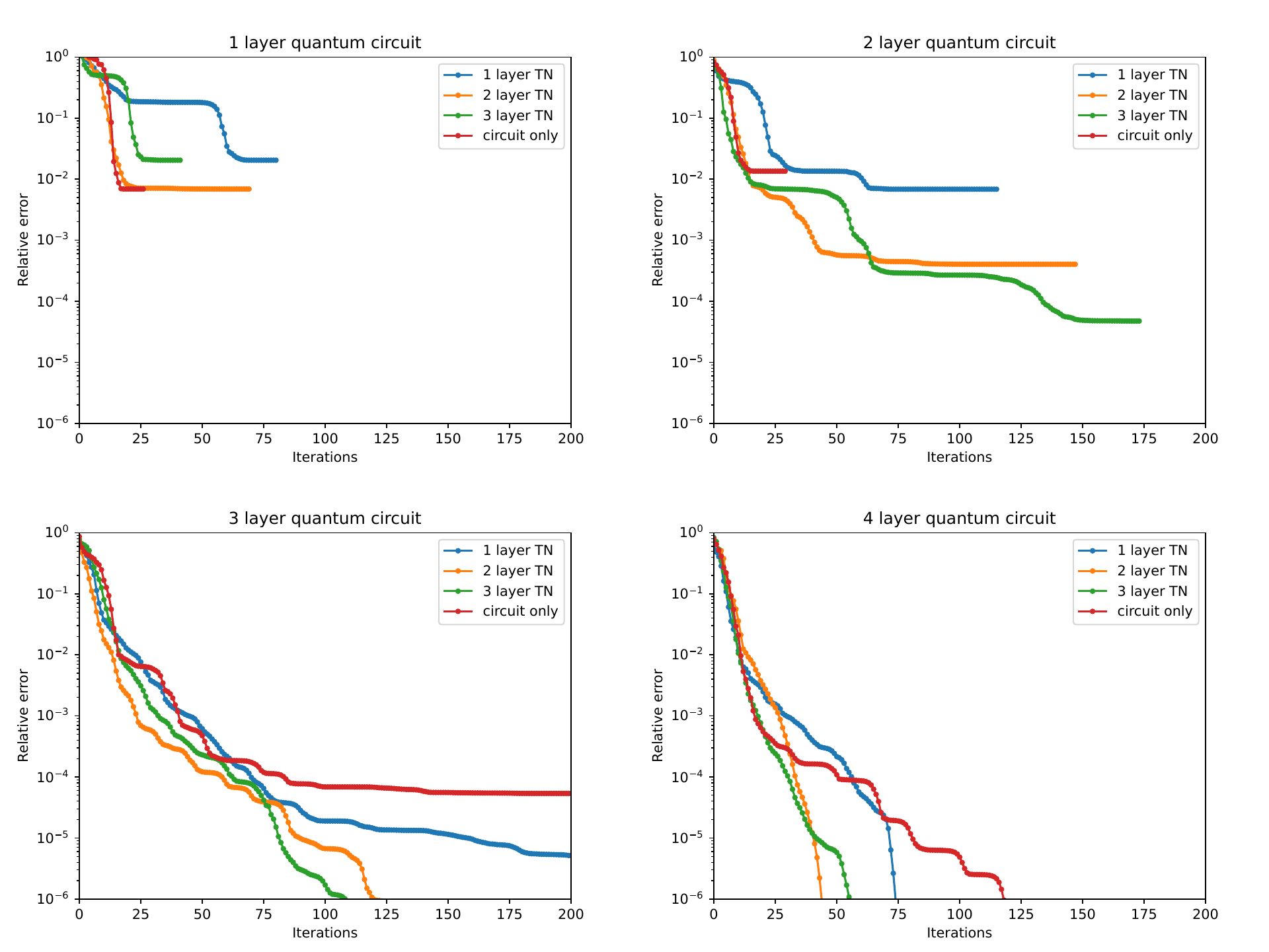}
    \caption{Relative error of the ground state energy of the transverse-field Ising model as a function of iterations in the VQE optimization procedure. The number of layers in the quantum circuits increases from 1 to 4 reading the plots from left to right. All tensor networks are constructed from tensors with three parameters. Lines which terminate before 200 iterations do so when the optimizer is unable to update the parameters due to a vanishing gradient.}
    \label{fig:many_parameter_results}
\end{figure}

The efficiency and performance of the suggested TN-VQE was compared to a classical VQE algorithm.
Two TN parameterizations were considered: the first one using a single parameter (Equation (\ref{eq:single_parameter_matrix})), and the other using three parameters (Equation (\ref{eq:arbitrary_matrix})).
Both the number of layers of the circuit and those of the TN were varied. The selected problem to solve was the 1D transverse-field Ising model with four sites.
\\
The results are presented in Figure \ref{fig:single_parameter_results}, for the single parameter TN, and Figure \ref{fig:many_parameter_results} for the multiple parameter case.

In both cases, the observed behaviour is similar. For a single layer quantum circuit, the optimization does not converge to the correct answer and the TN is unable to improve the result in any way, regardless of the number of layers in the TN. For a two layer quantum circuit, the optimization also does not converge to the correct answer, although the TN is occasionally able to improve on the result.
For example, with the three parameter two layer TN, the error in the answer is reduced by almost three orders of magnitude when compared to the circuit without a TN.
For a three layer quantum circuit, the circuit without a TN is still unable to converge to the correct answer, but in multiple instances, the TN is able to reduce the relative error to below $10^{-6}$.
For a four layer quantum circuit, both the TN-VQE and only VQE cases are able to converge to the correct answer.
In many cases the TN-VQE configurations are able to converge with fewer iterations, as can be seen from the bottom right plot of both Figures \ref{fig:single_parameter_results} and \ref{fig:many_parameter_results}.
Although not displayed in these plots, this behaviour also holds true for QCs with more than four layers.

\begin{figure}[htbp!]
  \centering
    \includegraphics[scale=0.29]{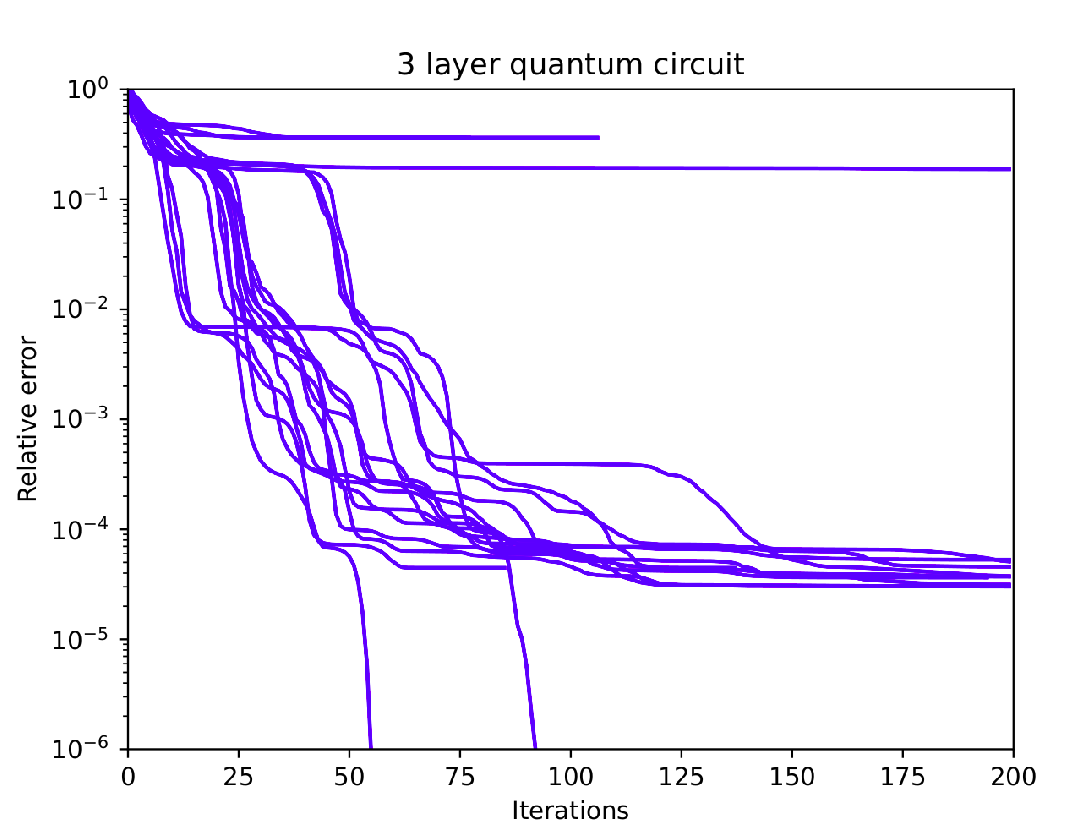}
    \caption{Performance of the randomly initialized PQC with 3 layers over 20 runs with a maximum of 200 iterations per optimization loop.}
    \label{fig:typical_behaviour_1}
\end{figure}

\begin{figure}[htbp!]
    \centering
    \includegraphics[scale=0.64]{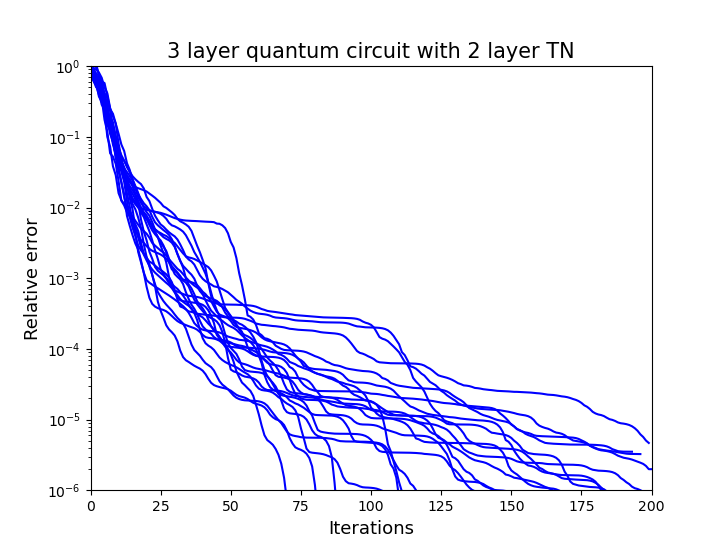}
    \caption{Perfomance of the randomly initialized PQC with 3 layers quantum circuit and 2 layers TN over 20 runs with a maximum of 200 iterations per optimization loop.}
    \label{fig:typical_behaviour_2}
\end{figure}

In Figure \ref{fig:typical_behaviour_1} and Figure \ref{fig:typical_behaviour_2} it is shown that the behaviour observed in Figures \ref{fig:single_parameter_results} and \ref{fig:many_parameter_results} is the typical efficiency to be expected from PQCs. Thus, stating more clearly that convergence towards the right solution is more likely with the TN-VQE approach than with randomly initialized circuits.

\begin{figure}[htbp!]
    \centering
    \includegraphics[scale=0.8]{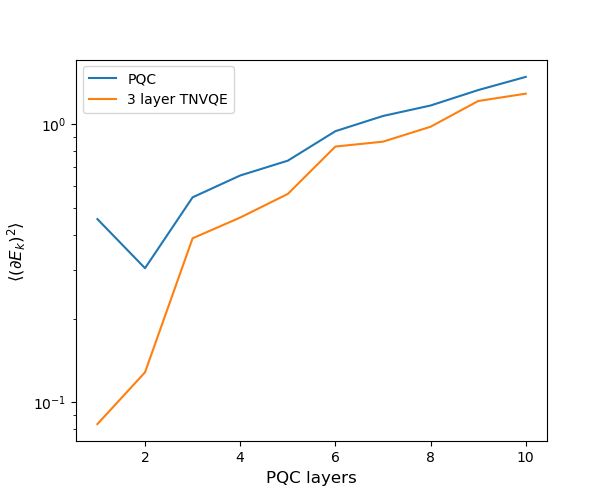}
    \caption{Variance of the gradients magnitude of the randomly initialized circuit (PQC) versus the joint TN-VQE with a 3 layer TN. The variance is calculated following Equation \ref{eq:variance}, where the mean is obtained over 200 iterations.}
    \label{fig:gradient_magnitudes}
\end{figure}

In terms of barren plateaus in the optimization landscape, it can be discussed that the 2 and 3 layer diagrams seem to show the optimization getting stagnant at a certain value; thus, suggesting 'barrenness'. To further investigate this behaviour, random samples of parameters in the landscape were taken, and the gradient variance was monitored. The variance was estimated following the expression from \cite{2018NatCo...9.4812M}:
\begin{equation}\label{eq:variance}
    Var{} [\partial_k E] = \langle (E_k)^2 \rangle
\end{equation}
where $k$ are the chosen parameters to parameterize the quantum circuit ($\phi$), or both the circuit and TN ($\theta$) in the TN-VQE.
\\
If there was to be a sharp decrease in variance, it would point towards the existence of a barren plateau. Based on the fact that deeper PQCs have shown to be prone to barren plateaus, the gradients were monitored when increasing the layers of the circuit, and averaged over 200 iterations. However, the variance did not seem to drop significantly with the increase of layers. This could be due to the scale of our circuit, or perhaps the small number of qubits, not allowing to spot certain trends. It could also be that the apparent 'barrenness' is just a minima and not necessarily a plateau.
\\
From the study of the variance, some conclusions can still be inferred about the efficiency of the TN-VQE method versus PQC. As seen in Figure \ref{fig:gradient_magnitudes}, the gradients' magnitudes are smaller overall for the TN-VQE method than for the randomly initialized circuit. This suggests that the parameters are being updated more efficiently in the joint method.

It might mean that the optimization algorithm converges to a minimum with smaller steps, indicating a more direct path to the optimal solution. A smoother cost function landscape often results in smaller gradient magnitudes. If the landscape is relatively flat around the optimal solution, then the algorithm does not need to take large steps to find the minimum. This could be an indication that the cost function is well-behaved and less prone to getting stuck in local minima for the TN-VQE, which is supported by the results in Figures \ref{fig:single_parameter_results} and \ref{fig:many_parameter_results}. Also, smaller gradient magnitudes might be less sensitive to the noise of fluctuations in the quantum simulation, hence being more advantageous for their practical implementation.

\section{Conclusion}\label{sec:conclusion}
In this work a many-body Hamiltonian was simulated by applying TNs to perform joint optimization with a VQE algorithm in order to obtain the ground state values for a 1D transverse field Ising Hamiltonian.
The Hamiltonian was constructed and parameterized with an unitary MPO. The larger the amount of parameters within $\theta$, the more efficient the performance of the TN-VQE was observed. However, it must be noted that even with a single parameter the performance of the TN-VQE still surpassed that of the VQE. This is of special interest given that the less parameters to optimize, also the less computational resources are being consumed.
\\
By merging the strength of the classical TN methods with the quantum VQE optimization scheme, we have demonstrated a higher accuracy and efficiency for obtaining the correct solution. This improvement in performance was achieved without varying the circuit, meaning it was possible to keep a shallow circuit depth yet obtain better results. This shows TN methods as a great tool for the implementation of PQCs on current NISQ devices.
\\
It still remains an open question if they can completely evade the barren plateaus present in the optimization landscape.

\bibliography{./references.bib}

\end{document}